\input epsf
\documentstyle[pra,aps, twocolumn, epsf ]{revtex}
\begin{document} 
\preprint{}
\draft
\twocolumn[\hsize\textwidth\columnwidth\hsize\csname @twocolumnfalse\endcsname
\title{ Superrevivals in the quantum dynamics of a particle confined
in a finite square well potential} 
\author{Anu Venugopalan\cite{email1} and  G. S. Agarwal \cite{email} 
} \address{
Physical Research Laboratory, Navrangpura\\ Ahmedabad, ,
INDIA -380 009} 
\maketitle 

\begin{center}
\end{center}

\begin{abstract} 
We examine the  revival features in  wave packet dynamics of a particle
confined in a  finite square well potential. 
The possibility of tunneling modifies the revival pattern as compared to
an infinite square well potential. We study the dependence of  the revival 
times on the depth of the square well and predict the existence of 
superrevivals. The nature of these superrevivals is compared 
with similar features seen in the dynamics of wavepackets in an
anharmonic
oscillator potential. 

\end{abstract}

]

\section{Introduction}

The wavepacket in quantum mechanics is often viewed as the most 
`nearly classical' state and is known to exhibit many striking classical
properties. However, its inherent quantum nature also causes it to exhibit
many quantum mechanical features. It is known\cite{ps,arz} that in certain
nonlinear quantum systems the dynamics of the wave packet incorporates quantum
interference effects which cause it to undergo a sequence of collapses and
revivals and in the course of its evolution the wavepacket periodically breaks
up and reconstitutes its original form. At intermediate times the wavepacket
gathers into a series of subsidiary wavepackets called fractional 
revivals\cite{ap}. 
Quantum revivals are often a 
manifestation of the fact that the time evolution of the wavepacket is
driven by a discrete eigenvalue spectrum and revival features 
depend on the way the eigenenergies of the quantum system  
depend on the quantum number, $n$\cite{ap,bkp}.
Recently there has been a lot of interest in the
theoretical and experimental study of quantum revivals in a variety of
nonlinear systems like that of Rydberg atom wavepackets\cite{arz}, molecular
wavepackets\cite{mol}, wave packets in semiconductor quantum wells 
\cite{qwell} etc.
Most experiments dealing with quantum revivals
till
date focus on atomic\cite{arz} and molecular systems\cite{mol}, 
photon cavity systems\cite{photon}
and ions in traps\cite{wine}.
The simplest class of 
systems for which one would see fractional and full revivals are those for
which the energy spectrum goes as $n^2$, e.g., the infinite square well
potential and the rigid rotator\cite{bkp}. For systems with a nonquadratic 
dependence on $n$, one can see a new sequence of collapse and revivals 
which are different from the usual fractional revivals. Often these
sequences culminate with the wavepacket resembling its original form more
closely. These are superrevivals\cite{bkp,super}. The revival patterns of Rydberg
wavepackets, where the energy spectrum is nonquadratic in $n$, 
have been seen to exhibit such superrevivals\cite{super}.

An ideal model system to illustrate the
fractional revival phenomenon is the infinite square well potential since 
simple 
analytical forms of the eigenvalues
and the eigenfunctions allow for an easy analysis of the time evolution of
any initial state. This system has been recently studied in great detail in
the context of fractional revivals\cite{stroud,berry} as well as for the
space-time structures\cite{sch} that appear in the dynamical evolution. 
An experimental realization 
of the predictions of the fractional revival phenomenon in the infinite well
is most likely to be in semiconductor quantum well systems. In reality, 
however, it is rather impossible to find a physical system that creates
a truly infinite confining potential. It becomes crucial, therefore, to
to study the problem of revivals in more realistic, {\em physical} systems 
which are better described by finite well potentials rather than infinite
potentials. The possibility of the occurrence of both revivals and 
superrevivals in a finite potential well has been pointed out 
before\cite{stroud} though no explicit study has been done till now.  
From a more fundamental point of view, the motivation
to study  the finite well system is
to get a greater
insight into understanding the
`classical limit'. During the course of its evolution an initially
localized wavepacket appears at certain
times as a linear superposition of spatially separated copies of
itself, i.e., in Schr\"{o}dinger cat-like states.
The tremendous
progress in experiments
involving semiconductor systems, e.g, the observation of
quantum beats in quantum wells\cite{qwell} and Bloch oscillations in semiconductor
superlattices\cite{wei} increases the prospects of generating, detecting
and studying such cat-like states among other features like revivals, thus
providing
further motivations to get a better theoretical
understanding of these systems.

In this paper we focus on the revival features of wavepackets in {\em finite} 
square well potentials and compare it with the corresponding case for 
the infinite well. For our purpose we have only concerned ourselves with 
the bound eigenstates of the finite square well potential. 
According to quantum mechanics, the wavefunction has 
nonvanishing values in the classically forbidden regions thus giving
a nonvanishing probability for the particle being outside the well. One
thus expects a  difference in the wavepacket dynamics as compared to 
the infinite well case.   
We show that the 
existence of the possibility of tunneling modifies the revival pattern as
compared to the infinite square well potential. In particular it allows for
revivals {\em and} superrevivals. We show that usual revival
times are now longer compared to that of the infinite well and depend 
on the well depth. 
The paper is organized as follows. In Section II
we begin by introducing the concept of fractional revivals and review 
briefly the finite well problem. We then present our 
numerical results and confirm the approximate formula of 
Barker {\em et al.}\cite{bark} 
from which we deduce an approximate analytical expression for  
revival times in finite square well potentials. In Section III we present 
another key result of this paper which is the existence of superrevivals 
during the time evolution of an initial Gaussian wavepacket. 
The absence of an exact analytical expression for the energy spectrum
for the finite potential well makes it difficult to give an estimate of the
superrevival times. However, we get some insight into the nature of 
superrevivals in the well by comparing it with superrevivals in the dynamics
of Gaussian wavepackets in an anharmonic oscillator
potential for which we have the analytical results. 
Finally, we summarize our results in Section IV. 

\section{Dependence of revival times on well-strength}

Consider the time evolution of a particle initially in state $\psi(x,0)$ in
a potential:
\begin{equation}
\psi(x,t)=\sum_{n} c_{n}\phi_{n}(x)e^{-iE_{n}t/\hbar},
\end{equation}
where $n$ is the quantum number and $\phi_{n}(x)$ and $E_{n}$ are the
energy eigenstates and corresponding eigenvalues. The coefficients $c_{n}$
are given in terms of the  initial wavefunction by 
$c_{n}$=$\langle\phi_{n}(x)|\psi(x,0)\rangle$. In general the superposition
(1) may also contain  continuum states for which $n$ would be a continuous 
index and the sum would be replaced by an integral. Here, however, we
concern ourselves with only superpositions of bound states assuming 
negligible continuum contributions. Also, one assumes that the expansion
(1) is strongly weighted around a mean value, $\bar{n}$. Both the 
assumptions above are reasonable in, e.g., the experimental situation when 
a localized wavepacket is produced using a short laser pulse\cite{ps}.
If one assumes that 
the weighting probabilities $|c_{n}|^{2}$ are strongly centered around a mean
value $\bar{n}$, one can expand the energy in a Taylor series in $n$ around 
$\bar{n}$ as:

\begin{eqnarray}
E_{n}&=& E_{\bar{n}} + E_{\bar{n}}'(n-\bar{n}) + 
\frac{1}{2}E_{\bar{n}}''(n-\bar{n})^{2} \\ \nonumber
&& + \frac{1}{6}E_{\bar{n}}'''
(n-\bar{n})^{3} + ....
\end{eqnarray}
where the primes denote derivatives with respect to $n$. From (2) one
can identify the time scales,
\begin{equation}
T_{cl}= \frac{2\pi\hbar}{|E_{\bar{n}}'|}, \hspace{.2in} T_{rv}=\frac{2\pi\hbar}
{\frac{1}{2}|E_{\bar{n}}''|}, \hspace{.2in} T_{sr}=\frac{2\pi\hbar}
{\frac{1}{6}|E_{\bar{n}}'''|},  ...
\end{equation}
etc. which are generally termed in the literature as the classical, revival, 
superrevival times and so on\cite{bkp}. 
One can easily see that re-writing the time 
evolution (1) 
in terms of these time scales shows how they govern the time evolution of 
$\psi(x,0)$. The time scales in turn are controlled by the dependence of the
energy on the quantum number $n$.  For the simple case of the infinite well,
the quantized energy levels are exactly quadratic in $n$:
\begin{equation}
E_{n}=
\frac{\pi^{2}\hbar^{2}n^{2}}{2mL^{2}}, 
\end{equation}
where $m$ is the mass of the particle and $L$ is the length of the well.
 Corresponding to this, one has the time scales 
$T_{cl}=2mL^{2}/\pi\hbar\bar{n} $ ,  $T_{rv}=4mL^{2}/\pi \hbar$ 
while $T_{sr}=\infty$. The time evolution
(1) can be re-written as:
\begin{equation}
\psi(x,t)=\sum_{n} c_{n}\phi_{n}(x)e^{-2\pi i(t/T_{rv})n^{2}}.
\end{equation}
The expansion (5) includes both odd and even parity states.
It is easy to see that the wavefunction regains its original form
i.e., shows full revivals, whenever $t$ equals some multiple of $T_{rv}$,
and shows fractional revivals whenever $t$ is equal to some rational
fraction of $T_{rv}$, e.g, $T_{rv}/4$, but no higher order effects
like superrevivals are seen in the dynamics of the wavepacket for the 
infinite square well potential\cite{stroud}. 

We now turn to the problem of the finite square well potential which is the 
focus of this paper.
The one-dimensional finite square well potential for a well of length $L$
can be described as:
\begin{eqnarray}
V(x)&=&0, \hspace{.4in} |x| \leq \frac{L}{2} \\ 
V(x)&=&V_{0},\hspace{.3in} |x| > \frac{L}{2} \nonumber.
\end{eqnarray}
The system thus has three distinct regions and the 
solutions to the Schr\"{o}dinger equation gives us the energy eigenstates and 
eigenvalues. The even parity solutions for the three regions are:
\begin{eqnarray}
\phi_{I}(\bar{x})&=&A \hspace{.1in}e^{\beta_{n}}\hspace{.1in} 
\cos{\alpha_{n}}\hspace{.1in} e^{2\beta_{n}\bar{x}}, \\  \nonumber
\phi_{II}(\bar{x})&=&A \hspace{.1in} \cos{2\alpha_{n}\bar{x}}, \\ \nonumber
\phi_{III}(\bar{x})&=&A \hspace{.1in}e^{\beta_{n}}\hspace{.1in}
\cos{\alpha_{n}}\hspace{.1in} e^{-2\beta_{n}\bar{x}}, 
\end{eqnarray}
where the eigenvalues are evaluated by solving the transcendental equation
\begin{equation}
\alpha_{n}\tan{\alpha_{n}}=\beta_{n},
\end{equation}
while the odd parity solutions are given by:
\begin{eqnarray}
\phi_{I}(\bar{x})&=&-A \hspace{.1in}e^{\beta_{n}}\hspace{.1in}
\sin{\alpha_{n}}\hspace{.1in} e^{2\beta_{n}\bar{x}}, \\  \nonumber
\phi_{II}(\bar{x})&=&A \hspace{.1in} \sin{2\alpha_{n}\bar{x}}, \\ \nonumber
\phi_{III}(\bar{x})&=&A \hspace{.1in}e^{\beta_{n}}\hspace{.1in}
\sin{\alpha_{n}}\hspace{.1in} e^{-2\beta_{n}\bar{x}},
\end{eqnarray}
with eigenvalues given by the transcendental equation
\begin{equation}
\alpha_{n}\cot{\alpha_{n}}=-\beta_{n}.
\end{equation}
Here $\bar{x}=x/L$ where $L$ is 
the length of the square well, $\alpha_{n}=\sqrt{mEL^{2}/2\hbar^{2}}$ 
and
$\beta_{n}=\sqrt{m(V_{0}-E)L^{2} /2\hbar^{2}}$, with $E$ the energy 
$(E < V_{0})$ and
$m$ the mass of the
particle and the normalization constant $A$ is given as:
\begin{equation}
A=\sqrt{\frac{2}{1+\frac{1}{\beta_{n}}}}.
\end{equation}
It is clear from (7) and (9) that the wavefunction has nonvanishing values
in both the `classically forbidden' regions $I$ and $III$. The quantum
mechanical probability for the particle to be somewhere in regions $I$ and
$III$ is, therefore, nonvanishing. It is clear, though, that as one goes
away from the boundaries, the probability density decreases rapidly to zero. 
Thus the particle with $E < V_{0}$ cannot really escape to infinitely long
distances but stays `bound' to the well. 
One can define 
\begin{equation}
\epsilon=\sqrt{\frac{mV_{0}L^{2}}{2\hbar^{2}}}
\end{equation}
as the `well-strength'\cite{bark}. 
A finite well of well strength $\epsilon$ would
contain a finite number of bound states, $N$, where  
$N \sim \frac{2\epsilon}{\pi} +1$. The time evolution of any given initial
states can be expressed in terms of these eigenstates and eigenvalues, 
\begin{equation}
\psi(x,t)=\sum_{n}^{N} c_{n}\phi_{n}(x)\exp{(-\frac{8i\alpha_{n}^{2}\tau}
{\pi})},
\end{equation}
where $\tau=t/T_{rv}$ is the time scaled in terms of the revival time $T_{rv}=
4mL^{2}/\pi\hbar$
of the infinite well potential. The expansion (13) contains both odd
and even parity states. 
For our purpose we examine the quantum dynamics of an initial Gaussian 
wavepacket with mean position $x_{0}$ and zero mean momentum:
\begin{equation}
\psi(x,0)=B \exp{ \Big (-\frac{(\bar{x}-x_{0})^{2}}{2\sigma^{2}}\Big )},
\end{equation} 
where $B$ is the normalization constant. The time evolution involves the 
energy eigenvalues, $\alpha_{n}s$  and the corresponding eigenfunctions 
which are obtained by 
solving the transcendental equations
(8) and (10) numerically. 
A Gaussian wavepacket of zero mean momentum can be faithfully
constructed by a superposition of these bound states ( the sum of the
coefficients, $\sum|c_{n}|^2 \approx 1$ ). 

Our numerical simulations for the finite well show the presence of 
revivals in the dynamics
of the initial Gaussian wavepacket (14) similar to that seen in the case 
of the infinite well potential for short times. 
The revival times, however, are in 
general longer than that of the infinite well and depend on the well strength
(depth), $\epsilon$, of the finite well. 
Barker {\em et al.}\cite{bark} have shown via a first order Taylor series expansion
of the transcendental equations (8) and (10) for the eigenvalues that the 
energy levels of a finite well of length $L$ and 
well strength $\epsilon$ can be approximated as:
\begin{equation}
E_{n}'=
\frac{\pi^{2}\hbar^{2}n^{2}}{2mL^{2}} \Big 
( \frac{\epsilon}{1+\epsilon} \Big )^2.
\end{equation}
One can see that this is equivalent to the energy levels for an {\em infinite}
well but with the larger length, 
\begin{equation}
L'=L\Big(1+ \frac{1}{\epsilon} \Big ).
\end{equation}
Corresponding to this approximate expression we can thus find an expression
for the approximate revival time, $T_{rv}'$, for the finite well in terms of
$T_{rv}$, the revival time for the infinite well:
\begin{equation}
T_{rv}'=T_{rv}\Big(1+ \frac{1}{\epsilon} \Big )^2.
\end{equation}
The formula of Barker {\em et al.} is more accurate for tightly bound eigenstates 
(deeper wells) than for weakly bound states (shallow wells)\cite{bark}

The revival features for the dynamics of any initial state can be 
understood by examining the absolute square of the autocorrelation function:
\begin{eqnarray}
|A(\tau)|^{2} &=& |\langle \psi(x,0)\psi(x,\tau)\rangle|^{2} 
=
|\sum_{n}|c_{n}|^2 
e^{-iE_{n}t/\hbar}|^{2} \\ \nonumber
&=& | \sum_{n}|c_{n}|^2  e^{-8i\alpha_{n}^{2}\tau/\pi}|^{2}.
\end{eqnarray}
Fig. 1 shows the square of the autocorrelation function, 
$| A(\tau)
|^2$, for an initial Gaussian wavepacket 
for three different $`$well-strengths', $\epsilon$, contrasted with that for
the infinite potential. 
One can see that for larger values of $\epsilon$, the revival time approaches 
that of the infinite well
($T_{rv}' \rightarrow T_{rv}$). The agreement of the actual revival
times with the 
formula of Barker {\em et al.} 
gets more accurate as $\epsilon$ increases, as expected and as shown in 
Table I. Thus for short times the dynamics of the wavepacket in the finite 
well is similar to that in the infinite well with modified revival times
which depend on the depth of the well. For our simulations we work with the
values of the parameters for the initial Gaussian state and the well-strength,
$\epsilon$ such that $\sum|c_{n}|^2 \approx 1$ always. In Fig.1, note that the
detailed behaviour of the autocorrelation function for the 
infinite well (solid line) shows a symmetry. In contrast, the detailed 
behaviour of the  
autocorrelation function for the finite well (dashed line) 
shows an asymmetry especially
around the revival time as is quite evident from Fig.1(a) which corresponds
to the shallowest of the three wells, ($\epsilon=12$). As the well gets
deeper, the asymmetry decreases and as expected, the detailed behaviour 
begins to resemble closer to the infinite well as is obvious from the
`deeper' well shown in Fig.1(c). 
\begin{figure}
\begin{flushleft}
\leavevmode
\epsfysize=12.0 cm \epsfbox{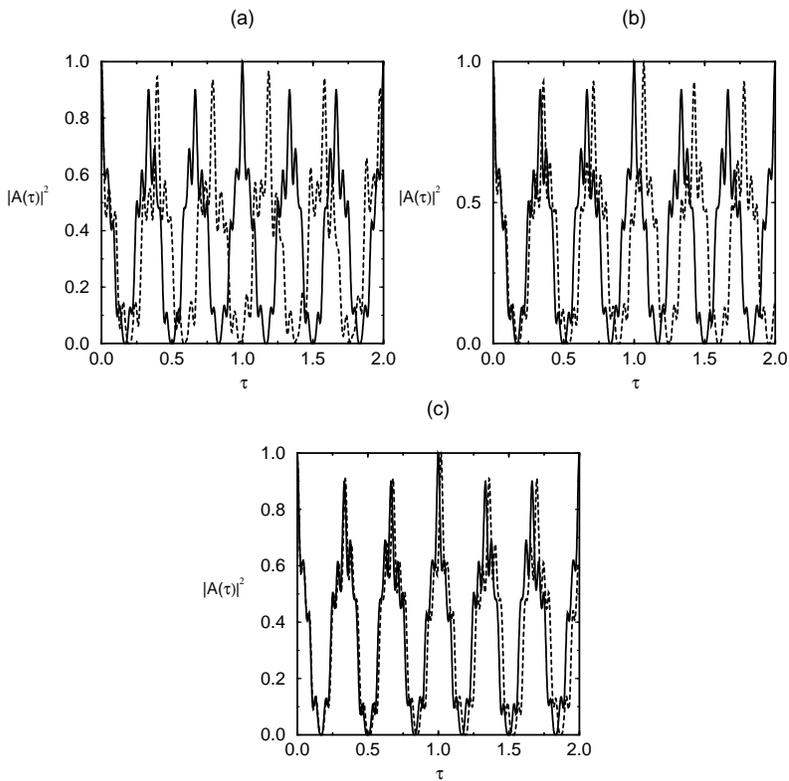}
\end{flushleft}
\vspace*{0.5cm}
\caption{ \bf Square of the autocorrelation function, $|A(\tau)|^2$ vs
the scaled time $\tau$ for an initial Gaussian wavepacket of width $L/10$
centered initially at $L/5$ for the infinite well (solid-line) and finite
well (dashed line) of well-strengths: (a) $\epsilon=12 (N\sim 8)$, (b)
$\epsilon=30(N\sim 20)$, (c)$\epsilon=100(N\sim 64 )$.
($\tau = t/T_{rv}, T_{rv}
= 4mL^{2}/\pi\hbar$)}
\end{figure}
\begin{figure}
\begin{center}
\begin{table}
\begin{tabular}{|c|c|c|c|}
{$\epsilon$} & {$T_{rva}$} & {$T_{rvb}$} & percentage error \\
(well-strength) & ($4mL^{2}/\pi\hbar$) & ($4mL^{2}/\pi\hbar$)& \\
 & & & \\ \hline
12  ($N \sim 8$) & 1.185 & 1.174 & 0.9$\%$  \\
& & & \\ \hline
 30  ($N \sim 20$)& 1.06777 & 1.06876 & .09$\%$ \\
& & & \\ \hline
 100  ($N \sim 64$) & 1.02009 & 1.02005 & .0039 $\%$ \\
& & & \\
\end{tabular}
\vspace*{0.3cm}
\caption{ \bf  Comparison of the
revival times, $T_{rva}$ with the revival times estimated from the
approximate formula of Barker {\em et al.}, $T_{rvb}$, for various
well-strengths, $\epsilon$. All times are scaled by $T_{rv} =
4mL^{2}/\pi\hbar$,
the revival
time for the infinite square well.
}
\end{table}
\end{center}
\end{figure}
\section{Superrevivals}

In the preceding section we examined the wavepacket dynamics and the structure 
of fractional revivals at short times, i.e., times of the order of or close to
$T_{rv}'$. Now we turn to the behaviour at longer times. 
A look at the square 
of the 
autocorrelation function, $|A(\tau)|^{2}$, for longer times shows that after
a few revival cycles, the wavepacket ceases to `reform' faithful to its original
form at $t=0$. The peaks in  $|A(\tau)|^{2}$ signifying full revivals
go down and then pick up again after a few cycles and the wavepacket is 
once again closer to its original form. This new sequence of revivals
is characterized by a longer revival time ($T_{sr}$).
This
behavior depends, as expected, on the depth of the finite well and is 
more obvious when we are dealing with shallow wells, i.e., for 
smaller values of $\epsilon$. For larger depths, $\epsilon$, the usual
revivals continue for more number of cycles before the peaks in 
$|A(\tau)|^{2}$ go 
down and pick up again. The superrevival times, thus, get longer 
and longer with
increase in the well depth. This is in agreement with our expectation that
the behavior in a deeper finite well should approach that for the 
infinite
well. 
It is interesting to note that while the usual revival times at short
times decrease with increasing well-strength, $\epsilon$, and approach
that of the infinite well, $T_{rv}=4mL^{2}/\pi\hbar$, the superrevival
times increase with increasing $\epsilon$ approaching $T_{sr}=\infty$ for
the infinite well. Since the finite well eigenenergies and eigenfunctions
have no simple analytical form, it is not straightforward to esthave no simple analytical form, it is not straightforward to estimate
these superrevival times for various well-strengths $\epsilon$.
Figs. 2, 3 and 4 show the square of the autocorrelation function,
$|A(\tau)|^{2}$,
vs $\tau$ for an initial Gaussian wavepacket for two different vvs $\tau$ for an initial Gaussian wavepacket for two different values of the
well-strength, $\epsilon$. The structure of the plot of the
square of the autocorrelation
function vs the time $\tau$ is a reflection of the initial state and depends
on the parameters like the width and the mean position of the inon the parameters like the width and the mean position of the initial Gaussian
wavepacket. The arrow in these figures indicates the first superrevival.
We have taken the case of an initial wavepacket
which has its mean
position at the center of the well ($x_{0}=0.0$) for Figs. 2 andposition at the center of the well ($x_{0}=0.0$) for Figs. 2 and 3 and
($x_{0}=0.2L$) for Fig. 4.
\begin{figure}
\begin{flushleft}
\leavevmode
\epsfysize=12.0 cm \epsfbox{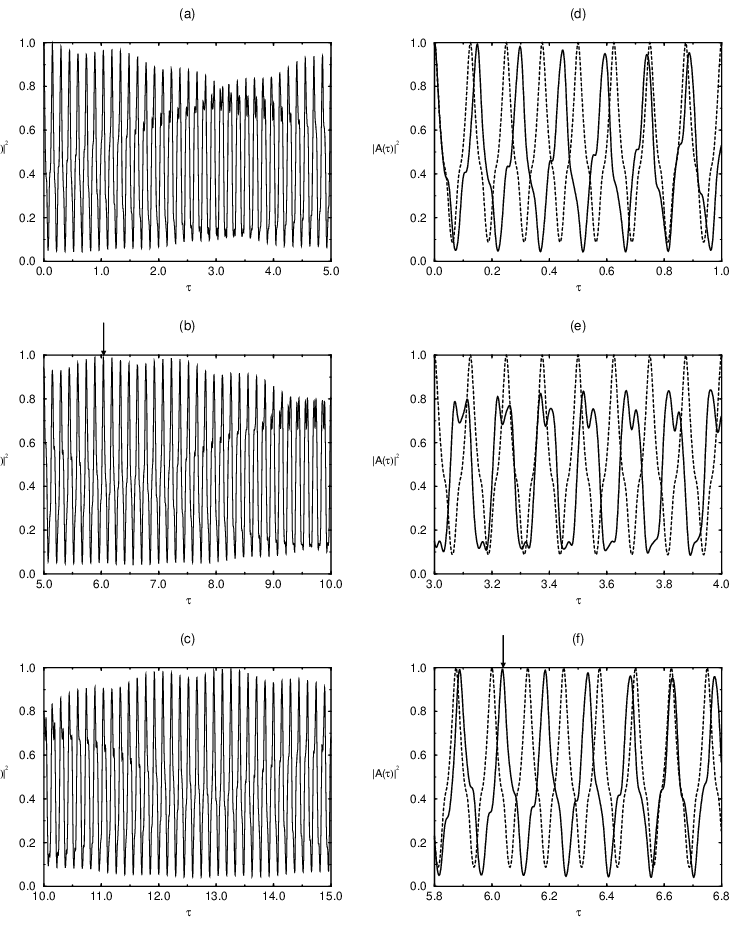}
\end{flushleft}
\caption{\bf Square of the autocorrelation function, $|A(\tau)|^2$ vs
the scaled time $\tau$
for an initial Gaussian wavepacket of width $L/10$
and with mean position at $\bar{x}=0.0$ for $\epsilon=12(N\sim 8)$;
(c) and (d) highlight the detailed behaviour at shorter times and
(e) highlights the detailed behaviour near the superrevival time(e) highlights the detailed behaviour near the superrevival time(e) highlights t
he detailed behaviour near the superrevival time.
The arrow marks the first superrevival.
The dashed curves correspond to the
corresponding plot for the infinite square well.
}
\end{figure}
One can see that there
are many revivals much before the characteristic time $T_{rv}'$ (e.g., there
are eight revivals in Fig.2(c) and Fig. 3(c) including the one at $T_{rv}'$).
Aronstein and Stroud\cite{stroud} have pointed out this feature in their
study of wavepacket dynamics in an infinite square well potential.
They show that for even parity states, such as the one where the
wavepacket is initially positioned at the center of the well 
($x_{0}=0.0$),
there are full revivals at multiples of $t=T_{rv}/8$. Similarly, odd
parity states show full revivals at multiples of times $t=T_{rv}/4$.
These features are obvious in the case of the dynamics in the infinite
square well potential when one looks at the time evolution of an initial
state in terms of the even and odd parity eigenstates:
\begin{eqnarray}
\psi(\bar{x},\tau)&=& \sum_{n(even)} \sqrt{2} e^{-2\pi in^{2}\tau} c_{n} 
\sin{n\pi
\bar{x}} \\ \nonumber
&&+ \sum_{n(odd)} \sqrt{2} e^{-2 \pi in^{2}\tau}c_{n} \cos{n\pi\bar{x}}.
\end{eqnarray}
Here $\tau=t/T_{rv}$ is the scaled time and $\bar{x}=x/L$.
For an initial even parity eigenstate $(\psi(\bar{x})=\psi(-\bar{x}))$,
only the second summation with the cosines contributes. 
It can be checked that at
$t=T_{rv}/8$, this sum can be written as:
\begin{eqnarray}
\psi(\bar{x},\tau)&=&\sum_{n(odd)} \sqrt{2}  e^{-2 \pi in^{2}\tau}c_{n} 
\cos{n\pi\bar{x}} \\ \nonumber 
&=& \sum_{m}\sqrt{2} \cos{(2m\pi\bar{x})} 
e^{-i \pi m(m+1)}e^{-\frac{i\pi}{4}} \\ \nonumber
&=& e^{-\frac{i\pi}{4}}\psi(\bar{x},0); \hspace*{0.5cm} m=0,1,2,..... 
\end{eqnarray}
At $t=T_{rv}/8$ the state is the same as what it was at $t=0$ except
for a constant phase factor.  Thus there will be full revivals at multiples
of $t=T_{rv}/8$. Similarly, for the odd parity states 
$(\psi(\bar{x})=-\psi(-\bar{x}))$ the contribution will only be from the
first summation containing the sines. At  $t=T_{rv}/4$, the time evolved state
can then be written as:
\begin{eqnarray}
\psi(\bar{x},\tau)&=&\sum_{n(even)} \sqrt{2}  e^{-2 \pi in^{2}\tau}c_{n}
\sin{n\pi\bar{x}} \\ \nonumber
&=&\sum_{m} \sqrt{2} \sin{(2m \pi \bar{x})} e^{-2 \pi i m^{2}} \\ \nonumber
&=& \psi(\bar{x},0) ; \hspace*{0.5cm} m=0,1,2,.....
\end{eqnarray}
Thus full revivals occur here at multiples of $t=T_{rv}/4$. 
For the case where the initial Gaussian has mean position $x_{0}=0.2L$ (a
state which does not have a definite parity),
there is a complete revival only at $T_{rv}$ though partial
revivals can be seen at multiples of $t=T_{rv}/3$ (Fig.4(c)).
These symmetry features  have an interesting analog
in the regeneration characteristics of a field $E(x)$ of wavelength $\lambda$ p
ropagating through a multimode planar
waveguide of width $b$\cite{jay}. If the field is symmetric in the transverse
dimension ($E(x)=E(-x)$), its regeneration length is $L=b^{2}/\lambda$. An
antisymmetric field regenerates at a distance $2L$ and  arbitrarantisymmetric field regenerates at a distance $2L$ and  arbitrary fields
reproduce after $8L$.
In the case of wavepacket dynamics in square well potentials, these
symmetry aspects are more clearly understood for the
infinite square well potential. However, it is not surprising
that we see similar features for wavepacket dynamics in the
finite well analyzed here.
\begin{figure}
\begin{center}
\leavevmode
\epsfysize=12.0 cm \epsfbox{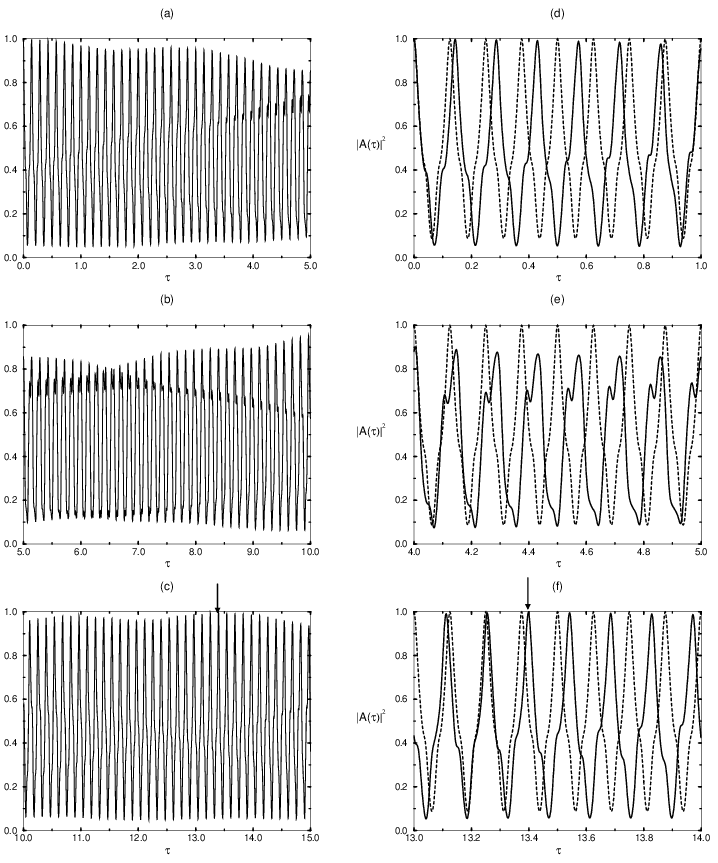}
\end{center}
\caption{\bf Square of the autocorrelation function, $|A(\tau)|^2$ vs
the scaled time $\tau$ for an initial Gaussian wavepacket of width $L/10$
and with mean position at $\bar{x}=0.0$ for
$\epsilon=15  (N\sim 10)$. (c) and (d)
highlight the detailed behaviour at shorter times and
(e) highlights the detailed behaviour near the superrevival time(e) highlights the detailed behaviour near the superrevival time(e) highlights t
he detailed behaviour near the superrevival time
The arrow marks the first superrevival. The dashed curves corresThe arrow marks
the first superrevival. The dashed curves correspond to the
corresponding plot for the infinite square well.
}
\end{figure}
Also, notice 
that for the initial Gaussian wavepacket centered at $x_{0}=0.0$, the 
autocorrelation function, $|A(\tau)|^{2}$ is never zero (Figs.2 and 3). On the 
other hand if we examine the case where  $x_{0}=0.2L$ (Fig.4), 
there will be times when $|A(\tau)|^{2}$  is zero 
and these times normally correspond to those instants when the 
wavepacket appears almost as a mirror reflection of itself and hence completely
uncorrelated with what it was at $t=0$. Local extrema of the autocorrelation
function, on the other hand, correspond to `fractional revivals', i.e, when the
wavepacket appears in a spatially separated superposition of 
replicas of itself
(cat-like states). 
As mentioned before, the
absence of a simple analytical form for the energy eigenvalues for the
finite well makes it difficult to find estimates for the superrevival
times. The
nature of these superrevivals and how often they occur would
also depend on various factors like the well-strength
and the initial position of the wavepacket.

\subsection*{{\bf Comparison with an anharmonic oscillator}}
As mentioned above, the lack of an analytical form for the
energy makes it difficult to estimate the superrevival times for the
finite well. However,
to get some insight, the revival and superrevival 
structure in wavepacket dynamics
of the finite well may be compared with the wavepacket dynamics in 
an anharmonic 
oscillator potential where one has the exact analytical expression for
the energy. We examine the hamiltonian with nonvanishing third order 
nonlinearity:
\begin{equation}
H=\mu_{1} (a^{\dagger}a)^{2} + \mu_{2} (a^{\dagger}a)^{3}.
\end{equation}
Here $a$ and $a^{\dagger}$ correspond to the annihilation and creation
operators, respectively. This system has been studied by Gantsog
et al\cite{qo}, though not in the context of quantum revivals. 
To make a comparison 
with the dynamics of the Gaussian wavepacket in the case
of the finite well, one can study the dynamics of 
initial coherent states and
squeezed states for the system described by (19).
We assume that $\mu_{2} << \mu_{1}$.
It can be easily checked 
that the revival times and superrevival times for the system (19) are
given by:
\begin{eqnarray}
t_{rv}' &=&  \frac{t_{rv}}{(1 + 3 \bar{n} \beta )}, \\ \nonumber
t_{sr} &=&t_{rv}/\beta,
\end{eqnarray}
where $t_{rv}=\frac{2 \pi\hbar}{\mu_{1}}$, is the revival time without the
third order term, i.e., with $\mu_{2}=0$ and $\beta=\frac{\mu_{2}}{\mu_{1}}$. 
$\bar{n}$ is the average photon number corresponding 
to the chosen initial state. 
Initial coherent and squeezed states can be expressed in
terms of the Fock states,  ${|n \rangle}$ , 
and one can study their time evolution.
The autocorrelation for an initial 
coherent state with amplitiude $\alpha$  can be easily shown to be
\begin{equation}
|A(\tau)|^{2}= |\sum_{n} \frac{|\alpha|^{n} e^{-\frac{|\alpha|^{2}}{2
}}
   }{n!} 
\exp{ \Big (-2\pi i \tau n^{2} - 2\pi i \beta \tau n^{3} \Big )} |^{2},
\end{equation}
while for an initial squeezed state, the autocorrelation function is given by
\begin{eqnarray}
|A(\tau)|^{2}&=&|\sum_{n}\frac{2 \sqrt{s}}{(s+1) 2^{n} n!} 
(\frac{s-1}{s+1})^{n} H_{n}^{2}\Big ( \frac{s \sqrt{2} \alpha}{\sqrt{s^2-1}}
\Big ) \\ \nonumber 
&&  \exp{\Big (-\frac{2 s \alpha^{2}}{s+1}-2\pi i \tau n^{2} - 
2\pi i \beta \tau n^{3} \Big )}|^{2}. 
\end{eqnarray}
Here $H_{n}$ denotes the $n$th order Hermite polynomial, $s$ is the
squeeze parameter\cite{nature} and $\tau=t/t_{rv}$ is the scaled time.
Although the system described by (19) is not quite  
the same as the finite well potential, one sees certain
similarities in the autocorrelation function in both cases. 
\begin{figure}
\begin{center}
\leavevmode
\epsfysize=12.0 cm \epsfbox{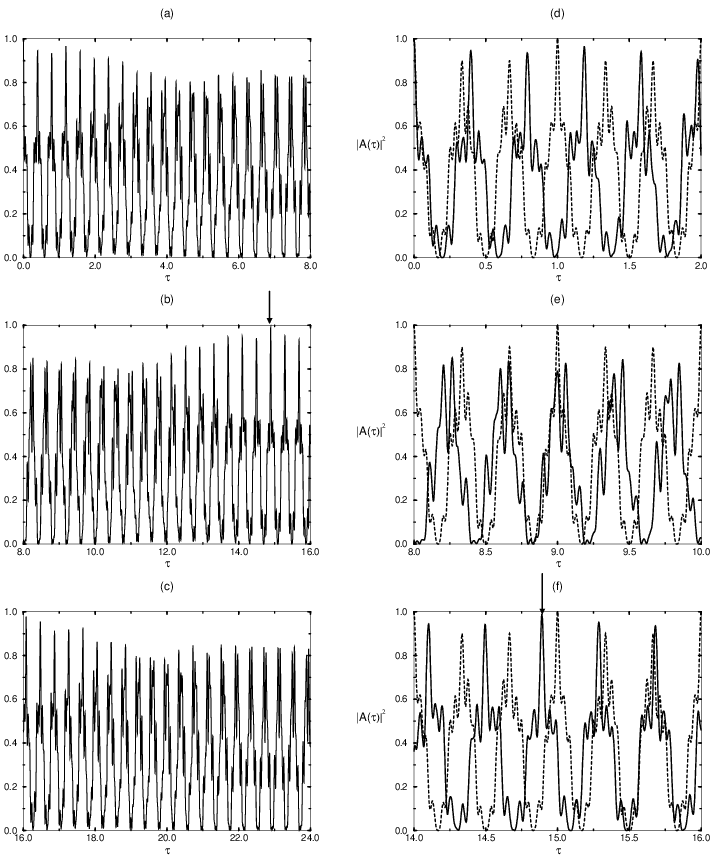}
\end{center}
\caption{\bf Square of the autocorrelation function, $|A(\tau)|^2$ vs
the scaled time $\tau$ for an initial Gaussian wavepacket of width $L/10$
and with mean position at $\bar{x}=
0.2L$ for $\epsilon=12(N\sim 8)$. (c) and (d)
highlight the detailed behaviour at shorter times and
(e) highlights the detailed behaviour near the superrevival time(e) highlights the detailed behaviour near the superrevival time
The arrow marks the first superrevival. The dashed curves correspond to the
corresponding plot for the infinite square well.
}
\end{figure}
The time evolution of an initial squeezed vacuum 
state ($\alpha=0)$ in the anharmonic oscillator 
potential (19) can be compared with
that of the narrow Gaussian wavepacket centered at $x_{0}=0.0$ 
in the finite quantum well.
Fig. 5 shows the autocorrelation function for the
dynamics of an initial squeezed vacuum in the the anharmonic potential
(19)  with squeeze parameter $s=10$, and $\beta=.002$ for short
times as well as for for times close to the superrevival times.
In both these cases one can see that the autocorrelation functions never
go to zero unlike the case when $x_{0}\neq 0.0$ for the packet in the
finite potential well and  $\alpha\neq0$ for the coherent state/squeezed state
in the anharmonic potential (results not shown for the anharmonic 
oscillator model). 
For this `symmetric case', there are many more revivals
at short times and superrevivals at longer times 
in both cases in contrast with 
the case when $x_{0}\neq 0.0$  and 
$\alpha\neq0$ when there are fewer revivals and superrevivals.
For the anharmonic oscillator the autocorrelation function 
also indicates the presence of fractional
superrevivals as seen in the finite well potential.
In particular one can compare the signature of the occurence 
of superrevivals
in both potentials by comparing 
the detailed structure of the autocorrelation
functions at times close to the superrevivals times. In both cases one
can see that at times nearing the superrevival times, the 
autocorrelation function begins to look more and more like 
the behaviour near $\tau=0$ and near the revivals times. Moreover, at
exactly these superrevival times the form 
of the autocorrelation function for the finite well appears very
similar to that of the infinite well and in the corresponding case
for the oscillator, the autocorrelation function at these times resembles
the behaviour for the case when $\mu_{2}=0$. 
At all other times away from
the revival and superrevival times, the autocorrelation functions look very
different. This is clear from Figs. 2, 3 and 4 where the detailed behaviour
of the autocorrelation function for the well
at short times and near the superrevivals times are
highlighted.
Thus the gross features regarding the nature of revivals and
superrevivals of
wavepackets in the finite
potential well are quite comparable with the dynamics of coherenpotential well are quite comparable with the dynamics of coherent and
squeezed states in a generic anharmonic oscillator potential (19).

\section {Conclusion}

In conclusion, we have studied the wavepacket dynamics in
a finite square well potential in the context of quantum revivals. We have
shown that for short times the revival patterns are similar to that
seen in the infinite well potentials but with a modified revival time which
is now dependent on the depth of the finite well. 
For deep enough wells,
the time scales
approach closer to that of the infinite well. For longer times, the
difference in the wavepacket dynamics as compared to the infinite
well potential is manifested by the appearance of superrevivals.
Superrevivals have till now been  predicted for the long time dynamics of
Rydberg
wavepackets\cite{super}. 
 
\begin{figure}
\begin{center}
\leavevmode
\epsfysize=12.0 cm \epsfbox{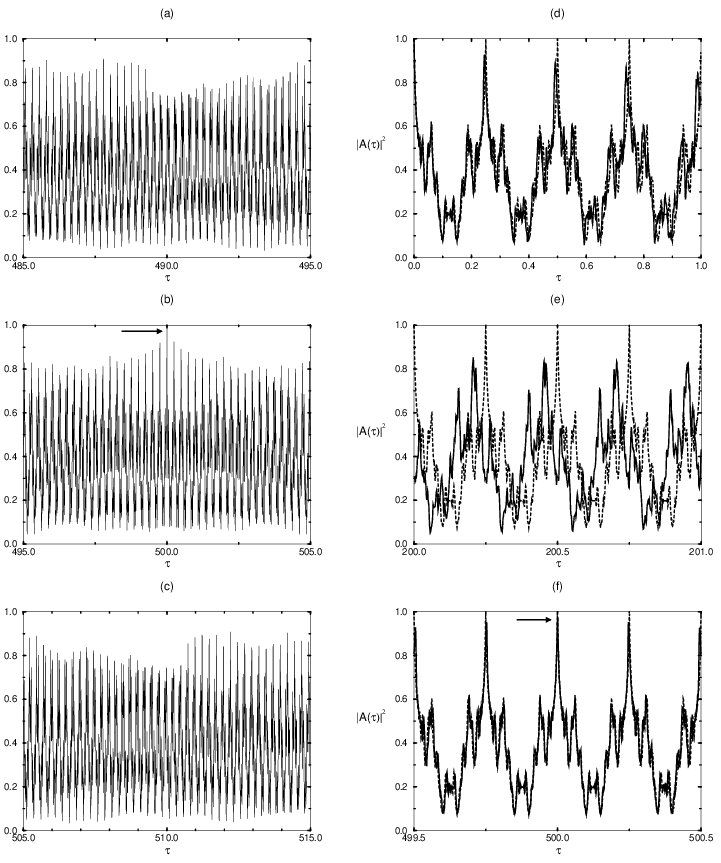}
\end{center}
\caption{\bf Square of the autocorrelation function, $|A(\tau)|^2$ vs
the scaled time $\tau$, $(\tau=t/t_{rv}, t_{rv}=2\pi\hbar/\mu_{1})$
for an initial squeezed vacuum with squeeze
parameter $s=10$ and $\beta=.002$ for the anharmonic oscillator
potential (19). (c) and (d)
highlight the detailed behaviour at shorter times and
(e) highlights the detailed behaviour near the superrevival time,
$T_{sr}=T_{rv}/\beta = 500T_{rv}$.
The arrow marks the first superrevival.
The dashed curves correspond to the
corresponding plot for the case when $\mu_{2}=0$
}
\end{figure}
The analysis in this paper predicts that wavepacket
dynamics in quantum well systems which can be more realistically
modelled as finite square well potentials rather than infinite ones,
will show superrevivals in addition to the usual revivals.

\vspace*{1cm}
One of us (GSA) is grateful to W. Schleich for useful discussions on the
dynamics of wavepackets.

\pagebreak


\begin{references}

\bibitem[\dagger]{email1} E-mail address:anu@prl.ernet.in

\bibitem[\ddagger]{email} E-mail address:gsa@prl.ernet.in 

\bibitem{ps} J. Parker and C. R. Stroud, Phys. Rev. Lett. {\bf 56}, 716 (1986);
Phys. Scr. {\bf T12}, 70 (1986)

\bibitem{arz} G. Alber, H. Ritsch and P. Zoller, Phys. Rev A {\bf 34}, 1058 (1986)

\bibitem{ap} I. Sh. Averbukh and N. F. Perelman, Phys. Lett. {\bf 139A}, 
449 (1989)

\bibitem{bkp} R. Bluhm ,V. A. Kostelecky and J. A. Porter, Am. J. Phys.
{\bf 64} (7), 944 (1996)

\bibitem{mol} B. M. Garraway and K-A Suominen, Rep. Prog. Physics {\bf 58},
365 (1995); W. S. Warren, H. Rabitz and M. Dahleh, Science {\bf 259}, 1581 (1993);
Marc. J. J. Vrakking, D.M. Villeneuve and Albert Stolow, Phys. Rev. A {\bf 54}, 37
(1996); I. Sh. Averbukh, Marc. J. J. Vrakking, D.M. Villeneuve and Albert
Stolow, Phys. Rev Lett. {\bf 77}, 3518 (1997).

\bibitem{qwell} K. Leo, J. Shah and E. O. Gobel, Phys. Rev. Lett. {\bf 66}, 201 (1991)

\bibitem{photon} G. Rempe, H. Walther and N. Klein, Phys. Rev. Lett. {\bf 58},
353 (1987); I. Sh. Averbukh, Phys. Rev. A {\bf 46}, R2205 (1992)

\bibitem{wine} D. M. Meekhof, C. Monroe, B. E. King, W, M. Itano and
D. J. Wineland, Phys. Rev. Lett. {\bf 76}, 1796 (1996).

\bibitem{super} R. Bluhm and V. A. Kostelecky, Phys. Rev A {\bf 50}, R4445 (1994);
R. Bluhm and V. A. Kostelecky, Phys. Lett. 200 A, {\bf 308} (1995)

\bibitem{stroud} D. L. Aronstein and C. R. Stroud, Phys. Rev. A {\bf 55} (6),
 4526 (1997)


\bibitem{berry} M. V. Berry, J. Phys. {\bf A 29}, 6617 (1996); M. V, Berry and
S. Klein, J. Mod. Optics {\bf 43}, 2139 (1996)

\bibitem{sch} F. Gossi$\beta$mann, J-M Rost and W. P. Schleich, J.Phys.
A Gen {\bf 30}, L277 (1997); P. Stifter, C. Leichtle, W.P. Schleich 
and J. Marklof, Z.
Naturf. {\bf 52}, 377 (1997); 
P. Stifter, W. E. Lamb, Jr and W. P. Schleich in {\em Frontiers of 
Quantum Optics and Laser Physics}, Eds. S. Y. Zhu, M. S. Zubairy and
M. O Scully (Springer, Singapore 1997), p 236 

\bibitem{wei} A. M. Weiner, J. Opt. Soc. Am. {\bf B 11}, 2480 (1994)

\bibitem{bark} B. I. Barker,G. H. Rayborn, J. W. Ioup and G.E. Ioup, Am.
J. Phys. {\bf 59}, 1038 (1991)

\bibitem{jay} J. Banerji, J. Opt. Soc. Am. {\bf B 14}, 2378 (1997);
J. Banerji, A. R. Davies and R. M. Jenkins, Appl. Opt. {\bf 36}, 1604 (1997).

\bibitem{qo} Ts. Gantsog and R.Tana\'{s}, Quantum Opt. {\bf 3}, 33 (1991).

\bibitem{nature} W. Schleich and J. A. Wheeler, Nature {\bf 326}, 574 (1987)

\end{references}
\end{document}